\documentclass[slac_one]{revtex4}
\usepackage{graphicx}
\usepackage{fancyhdr}
\begin{document}

\title{Monopoles and the 't Hooft tensor for generic gauge groups} %% Paper title goes here


\author{A. Di Giacomo}
\affiliation{Universita' di Pisa and INFN Sezione di Pisa,ITALY}
\author{L. Lepori}
\affiliation{ SISSA and INFN Sezione di Trieste,ITALY}
\author{F. Pucci}
\affiliation{Universita' di Firenze and INFN Sezione di
Firenze,ITALY}
\begin{abstract}

\end{abstract}

\maketitle We study monopoles and corresponding 't Hooft tensor in
a generic gauge theory. This issue is relevant to the
understanding of color confinement.
% in terms of.......
\thispagestyle{fancy}


\section{MOTIVATION} % Section title should be in all capitals.
The stringent experimental upper limits on the observation of free quarks in Nature
 indicate  that quarks are absolutely confined due to some symmetry.
 The deconfining transition is  a change of symmetry, i.e. an order disorder transition.
 Color is an exact symmetry. What can then be the extra symmetry responsible for confinement?
 The answer is provided by DUALITY.  Infrared modes exist with
topologically non trivial spatial boundary conditions (Homotopy)
which can be labelled by the eigenvalues of conserved topological
charges, i.e. in terms of a dual symmetry. Many examples of dual
symmetries are known in statistical mechanics \cite{1} and in
field theory\cite{2}. In (2+1) dimensions the homotopy group is
$\Pi _1$ and the excitations are vortices, in (3+1) dimensions the
homotopy group is  ${\Pi }_2$ and the excitations are monopoles.
This is equivalent to extend the formulation of the theory to a
spacetime with an arbitrary but finite number of line-like
singularities in each configuration (monopoles)\cite{3}.
\section{MONOPOLES}
 A prototype  is the 't Hooft-Polyakov monopole
\cite{4}\cite{5} in the $SU(2)$ gauge theory interacting with a
Higgs scalar \begin{math}\phi (r)=\phi^i(r) \sigma_i\end{math} in
the adjoint  representation. It is a static soliton solution made
stable by its non trivial homotopy.\newline In the ''hedgehog''
gauge at large $r$
\begin{equation}\phi^i \simeq \frac{r^i}{|\, \vec{r}\, |}\end{equation}
a mapping of the sphere \begin{math}S_2\end{math} at spatial
infinity on the group, with non trivial homotopy. In the unitary
gauge, where \begin{math}\hat \phi^i \equiv \frac{\phi^i}{| \phi\,
|}= \delta_3^i\, \sigma_3\end{math} is diagonal, a line
singularity appears starting from the location of the
monopole.\newline A gauge invariant field strength
\begin{math}F_{\mu \nu}\end{math} can be defined (t'Hooft
tensor), which coincides with the abelian field strength of the
residual symmetry  \begin{math}F^{3}_{\mu \nu }= \partial_{\mu}
A_{\nu}^{3} - \partial_{\nu} A_{\mu}^{3}\label{Abelian}\end{math}
 in the unitary gauge \cite{4}
\begin{equation}
F_{\mu\nu}=Tr(\hat{\phi}\, G_{\mu\nu})- \frac{i}{g}\, Tr
\left(\hat{\phi}\,
[D_{\mu}\hat{\phi},D_{\nu}\hat{\phi}]\right).\label{eq.4}\end{equation}
\vspace{0,2 cm}
 The monopole configuration has zero electric field
(\begin{math}F_{0 i}= 0 \end{math}). The magnetic field
\begin{math}H_i = \frac{1}{2}\, \epsilon_{ijk} F_{jk}\end{math} is
the field of a Dirac monopole of charge 2 \begin{equation} \vec{H}
= \frac{1}{g} \frac{\vec{r}}{4 \pi r^3}  \texttt{ + Dirac
String}\end{equation} In a compact formulation, like is lattice,
the Dirac string is invisible and a violation of Bianchi identity
occurs.\ A magnetic current can be defined as
\begin{equation}j_{\mu} =
\partial_{\nu}\widetilde{F}_{\mu \nu} \label{S}\end{equation} with
\begin{math} \widetilde{F}_{\mu \nu} = {1\over 2}{\epsilon}_{\mu \nu \rho \sigma} F_{\rho \sigma}\end{math}. A non
zero value of it signals the violation of Bianchi identities.
 The current  defined by eq.(\ref{S}) is identically
conserved due to the antisymmetry of  $ \widetilde{F}_{\mu \nu}$:
\begin{equation}\partial^{\mu} j_{\mu} = 0.\label{S1}\end{equation}
 This conservation law is the dual symmetry.

%The main feature of the definition  eq.(\ref{eq.4}) is that linear and bilinear
%terms in
%\begin{math}A_{\mu},A_{\nu}\end{math} cancel and one has
%identically
%\begin{equation}F_{\mu \nu} =  \frac{1}{2} Tr( \partial_{\mu}( \hat{\phi} A_{\nu} ) -
%\partial_{\mu}( \hat{\phi} A_{\nu} ) - \frac{i}{g}\, \hat{\phi} [ \partial_{\mu} \hat{\phi} ,
%\partial_{\nu} \hat{\phi} ] )\label{5}\end{equation}
% In the unitary gauge, where $\hat{\phi}=(0,0,1)$ and
%\begin{math}\partial_{\mu}\hat{\phi}=0\end{math}, it reduces to \begin{math}F^{3}_{\mu \nu }\end{math}.

The dual symmetry is well defined also in absence of Higgs
breaking. In a theory with no Higgs any operator  ${\phi}$ in the
adjoint representation can be used and monopoles will be located
at the zeroes of  ${\phi}$. The unitary gauges corresponding to
different choices of ${\phi}$  will differ by a gauge
transformation which is regular everywhere except for a finite
number of points. Creating a monopole means adding a singularity,
independent on the choice of ${\Phi}$ \cite{DiGiacomo:2002pe}.\\
 Recently some special
groups like $G_2$ and $F_4$ became of interest, since they have no
center and seem to confine \cite{6}, in contrast with the idea
that center vortices could be the configurations responsible for
confinement. However, for the group $G_2$ and $F_4$
%aggiunto qua
(unlike $SU(N)$ groups) it proves impossible to identify the
abelian field strength in the unitary gauge with a 't Hooft tensor
of the form of eq.(\ref{eq.4}),
%cambiato qui
because no solution exists for $\phi$ such that this equation is
true. Still there are monopoles and it is possible to define
magnetic conserved currents. However, the approach above has to be
modified for a more general construction of a 't Hooft like
tensor.

\section{MONOPOLES AND 'T HOOFT TENSOR FOR GENERIC GAUGE GROUP}
Let $G$ be the gauge group, which we shall assume to be compact
and simple. To define a monopole current we have to isolate an
$SU(2)$ subgroup, and break it to its third component, say $T_3$,
 by some "Higgs field" $\phi$  in the adjoint representation \cite{7}.\\
 Each of the roots of the Lie algebra identifies an $SU(2)$ subgroup.
 Since any root can be transformed to a simple root by a transformation of the group
 one can restrict the choice to the simple roots, which are as many as the rank $r$
 of the group and are represented by the circles in the Dynkin diagram of the Lie algebra.
 We are using here the language of any textbook on group representations, e.g.
 \cite{8}.
The "Higgs fields" $\phi^i$ are can be easily shown to be equal in
the unitary gauge to the fundamental weights $\mu^i$ corresponding
to each simple root $\vec \alpha ^i$, it transforms in the adjoint
representation of  $G$. The little group of $\phi^{\,i}$ is the
product of the $U(1)$ generated by $\phi^{\, i}$ times a group $H$
which has as Dynkin diagram a diagram (connected or not connected)
obtained by erasing from the diagram of $G$ the root $\alpha_i$
and the links which connect it to the rest (Levi subgroup), modulo
a subgroup of the center of $H\times U(1)$. Indeed $\phi=\mu^{\,
i}$ commutes with all the roots different from $\alpha_i$.
\newline There are $r$ independent dual charges $Q^a$
$(a=1,..,r)$, whose values are in one-to-one correspondence with
the homotopy classes of $\Pi _2$, and which are coupled to the
abelian field strengths along $\Phi^a$ in the unitary gauge, $F^a
_{\mu \nu}$.
\begin{equation}
j^a_{\nu} = \partial_{\mu} \tilde F^a_{\mu \nu}
\end{equation}
\begin{equation}
\partial_{\nu} j^a_{\nu} =0
\end{equation}
\begin{equation}
Q^a = \int d^3x j^0(\vec x, t)
\end{equation}
The field strengths  $F^a _{\mu \nu}$ can be written in a gauge
invariant form ('t Hooft  tensors). The result for a generic group
can be given in terms of the "Higgs"  fields $\hat \phi^a(x)$,
which coincide with the fundamental weights $\phi^a$ in the
unitary gauge,
%ho tolto che trasforma in aggiunta e l'ho messo
%all'inizio
and of the set of non zero values of  the numbers $\lambda^i _I=
(\vec \phi^i \vec \alpha)^2$, which are characteristic constants
of the group $G$. The explicit form is
 \begin{displaymath} F_{\mu \nu}^i = Tr ( \phi^i G_{\mu \nu} ) - \frac{i}{g}
\sum_{I} \frac{1}{\lambda_I^{i}  \,}\, \, Tr \left( \phi^i [D_{\mu} \phi^i, D_{\nu} \phi^i ] \right) +
\end{displaymath}
\begin{equation}
+\, \frac{i}{g} \sum_{I \neq J}\frac{1}{\lambda_{I}^{i}
\lambda_{J}^{i}}\, Tr \left( \phi^i [[D_{\mu} \phi^i, \phi^i],
[D_{\nu} \phi^i,\phi^i ]]\right) + ....\label{r1}\end{equation}
The sum runs on the different values of  $\lambda^i _I$ each taken
once. For $SU(N)$ groups $[\phi^i ,E_{\vec{\alpha}}]=( \vec{c}^{\,
\, i} \cdot \vec{\alpha} ) E_{\vec{\alpha}}$ and
% invece di where
$( \vec{c}^{\, \, i} \cdot \vec{\alpha}) = 0,\pm 1$, so that
 the 't Hooft tensor is the usual one
\begin{equation} F_{\mu \nu}^a = Tr ( \phi^a G_{\mu \nu} ) -
\frac{i}{g}Tr ( \phi^a [D_{\mu} \phi^a, D_{\nu} \phi^a ]
).\end{equation} For a generic group the projector is more
complicated and in principle depends on the root chosen \cite{7}.

 \section{CONCLUDING REMARKS}
\begin{itemize}
\item Experimental limits on the existence of free quarks indicate
that confinement is an absolute property due to some symmetry. The
deconfining transition is a change of symmetry. \item Color is an
exact symmetry: the extra symmetry responsible for confinement is
a dual symmetry related to spatial homotopy. Three dimensional
physical space implies that the extra symmetry is  generated by
magnetic charges. \item Higgs breaking of magnetic gauge symmetry
produces dual superconductivity and confinement. The transition to
the Coulomb phase is deconfinement. \item Magnetic conserved
current for a generic gauge group can be identified, and with them
the corresponding t'Hooft tensors. They label the dual degrees of
freedom.
\end{itemize}

\end{document}